\documentclass[11pt]{article}
\usepackage{amssymb,amsfonts,amsmath,amsthm}
\usepackage{subfigure}
\usepackage{multirow}
\usepackage{threeparttable}
\usepackage{graphicx}
\usepackage{subfigure}                                                                                          
\usepackage[bookmarks=false]{hyperref}
\usepackage{epstopdf}
\begin{document}


\title{Regulatory patterns in molecular interaction networks}

\author{David~Murrugarra\thanks{
           Corresponding author.  Address: 
           Virginia Bioinformatics Institute,
	   Virginia 
Polytechnic Institute and State University,
	   Blacksburg, VA~24061, U.S.A.,
	   Tel.:~(540)267-4112, Fax:~(540)231-2606} 
	   \and Reinhard Laubenbacher\\
Virginia Bioinformatics Institute and
	Department of Mathematics, \\
	Virginia 
Polytechnic Institute and State University, Blacksburg, VA} 

\date{}

\pagestyle{myheadings}
\markright{Regulatory patterns in molecular interaction networks}

\maketitle

\abstract{Understanding design principles of molecular interaction networks is an important
goal of molecular systems biology. Some insights have been gained into features of their network topology
through the discovery of graph theoretic patterns that constrain network dynamics. This paper contributes
to the identification of patterns in the mechanisms that govern network dynamics. 
The control of nodes in gene regulatory, signaling, and metabolic networks is governed by a 
variety of biochemical mechanisms, with inputs from other network nodes that act additively or synergistically. 
This paper focuses on a certain type of logical rule that appears 
frequently as a regulatory pattern. Within the context of the multistate discrete model paradigm, a rule type is introduced that
reduces to the concept of nested canalyzing function in the Boolean network case. It is shown that networks that employ
this type of multivalued logic exhibit more robust dynamics than random networks, 
with few attractors and short limit cycles. It is also
shown that the majority of regulatory functions in many published models of gene regulatory and signaling networks are
nested canalyzing.}



\section{Introduction}

Elucidating the large-scale graph structure of complex molecular interaction networks, from transcriptional
networks \cite{Bonneau} to protein-protein interaction networks \cite{Yu03102008} and metabolic network
\cite{Feist} is an important step toward an understanding of design principles of the cellular architecture. 
For instance, it has been shown that certain graph theoretic network patterns are much more prevalent in 
such networks than could be expected. See, e.g., \cite{Milo05032004}. The next step is to
understand cells as complex nonlinear dynamical systems. There now exist many dynamic models of 
gene regulatory, signaling, and metabolic pathways that provide snapshots of cellular dynamics 
using a range of modeling platforms. Many of these models represent
the interactions of different molecular species as logical rules of some type that describe the combinatorics of
how the species combine to regulate others; see, e.g., \cite{Naldi, Li06042004}. 
The logical rules of Boolean network models are an example of such a description, in which network states are
reduced to binary states, with a species either present or absent. It was shown in \cite{Kauffman2003, Kauffman2004}
that a special type of Boolean logical rule which appears frequently in published Boolean network models \cite{Harris}
exhibits robustness properties characteristic of molecular networks. These rules, so-called {\emph{nested canalyzing
functions}}, capture the spirit of Waddington's concept of canalyzation in gene regulation \cite{Waddington}. 
Several other classes of Boolean functions have also been investigated in the 
search for biologically meaningful rules to describe molecular interactions, including random functions \cite{Kauffman1}, hierarchical canalyzing function \cite{SzallasiLiang, Nikolajewa}, chain functions \cite{Gat-Viks03072003}, and unate functions \cite{Grefenstette}. 

In many cases the regulatory relationships are too complicated to be captured with Boolean logic, and more
general models have been developed to represent these. Common other discrete model types, in addition to 
Boolean networks, are so-called
logical models \cite{ThomasD'Ari}, Petri nets \cite{Steggles}, and agent-based models \cite{Pogson}. 
In \cite{Veliz-Cuba} and \cite{Hinkelmann}
it was shown that
many of these models can be translated into the rich and general mathematical framework of {\emph{polynomial
dynamical systems over a finite field}} $\mathbb{F}$. (Software to carry out this translation is available at 
{\tt http://dvd.vbi.vt.edu/cgi-bin/git/adam.pl}. Since the algebraic structure of $\mathbb F$ is not relevant for 
our purposes, we will consider a slightly more general setup. Let $x_1,\ldots , x_n$ be variables, which can take
values in finite sets $X_1,\ldots , X_n$, respectively.  Let $X = X_1\times\cdots\times X_n$ be the Cartesian product. 
A dynamical system
in the variables $x_1, \ldots , x_n$ is a function 
\begin{displaymath}
f = (f_1,\dots,f_n):X\rightarrow X
\end{displaymath}   
where each coordinate function $f_i$ is a function on a subset of $\{x_1,\dots,x_n\}$, and takes on values in $X_i$. 
Dynamics is generated by iteration of $f$. As an example, if $X_i = \{0, 1\}$, then each $f_i$ is a Boolean rule
and $f$ is a Boolean network.

Here, we use this very general framework to give a definition of the notion of {\emph{nested canalyzing
rule}}, which then applies to all different model types simultaneously. We show through extensive 
simulations that dynamical systems constructed from such rules as coordinate functions have 
important dynamic properties characteristic of molecular networks, namely very short limit cycles and
very few attractors, compared with the set of all possible functions. Furthermore, we show that many 
published models use logical interaction rules whose polynomial form is nested canalyzing, thereby providing
evidence that general nested canalyzing rules represent a frequently occurring pattern in molecular network
regulation. 

\section{Nested Canalyzing Rules}

Here we present the general definition of a nested canalyzing rule in variables $x_1,\ldots , x_n$ with
state space $X = X_1\times\cdots\times X_n$.

\medskip\noindent
{\bf Definition.}
Assume that each $X_i$ is totally ordered, that is, its elements can be arranged in linear increasing order.
In the Boolean case this could be $X_i = \{0 < 1\}$. 
Let $S_i\subset X_i, i = 1, \ldots , n,$ be subsets that satisfy the property that
each $S_i$ is a proper, nonempty subinterval of $X_i$, that is, every element of
$X_i$ that lies between two elements of $S_i$ in the chosen order
is also in $S_i$. Furthermore, we assume that the complement of each $S_i$ is also a 
subinterval, that is, each $S_i$ can be described by a threshold $s_i$, with all elements of $S_i$
either larger or smaller than $s_i$.

\begin{itemize}

 \item The function $f_i:X\rightarrow X_i$ is a {\it nested canalyzing rule} in the 
 variable order $x_{\sigma(1)},\dots,x_{\sigma(n)}$ with {\it canalyzing input sets} $S_1,\dots,S_n\subset X$ and {\it canalyzing output values} $b_1,\dots,b_n,b_{n+1}\in X_i$ with $b_n\neq b_{n+1}$ if it can be represented in the form: 

\begin{displaymath}
f(x_1,\dots,x_n)=\left\{
\begin{array}{l}
b_1\text{ if}\ x_{\sigma(1)}\in S_1\\
b_2\text{ if}\ x_{\sigma(1)}\notin S_1,x_{\sigma(2)}\in S_2\\
b_3\text{ if}\ x_{\sigma(1)}\notin S_1,x_{\sigma(2)}\notin S_2,x_{\sigma(3)}\in S_3\\
\vdots\\
b_n\text{ if}\ x_{\sigma(1)}\notin S_1,\dots,x_{\sigma(n)}\in S_n\\
b_{n+1}\text{ if}\ x_{\sigma(1)}\notin S_1,\dots,x_{\sigma(n)}\notin S_n
\end{array}\right.
\end{displaymath}

 \item The function $f_i:X\rightarrow X_i$ is a nested canalyzing function if it is a 
 nested canalyzing function in some variable order 
 $x_{\sigma(1)},\dots,x_{\sigma(n)}$ for some permutation $\sigma$ on $\{1,\dots,n\}$.
\end{itemize}

It is straightforward to verify that, if $X_i = \{0,1\}$ for all $i$, then we recover the definition in \cite{Kauffman2003} of a Boolean
nested canalyzing rule. As mentioned above, several important classes of multistate discrete models can be
represented in the form of a dynamical system $f:X\longrightarrow X$, so that the concept of a nested canalyzing rule
defined in this way has broad applicability. 

\section{The dynamics of nested canalyzing networks}

Aside from incorporating the biological concept of canalyzation, networks whose nodes are controlled by 
combinatorial logic expressed by nested canalyzing rules 
have dynamic properties resembling those of biological networks. In particular, they are robust,
due to the fact that they have a small number of attractors, which are therefore large. That is, perturbations are
more likely to remain in the same attractor. In addition, limit cycles tend to be very short, compared to random
networks, which implies that these networks have very regular behavior. 
We have performed extensive simulation experiments for this purpose, whose results we report here. 

\subsection{Computer experiments}

We have generated random network topologies with in-degree distribution $k$ ranging between 2 and 5, i.e., each node depends on at least two inputs and at most on five inputs. This assumption is not unrealistic and is based on the observation that
gene regulation networks are sparse \cite{Leclerc}.
For each network topology we have generated two discrete dynamical systems: one where the update rules are all nested canalyzing and the other where the update rules are randomly chosen (and non-degenerate,
in the sense that all inputs indicated in the network topology are realized).
 
\subsubsection{Attractor distribution for nested canalyzing networks versus random networks}

\begin{figure}[h]
  \begin{center}
        \includegraphics[width=\textwidth]{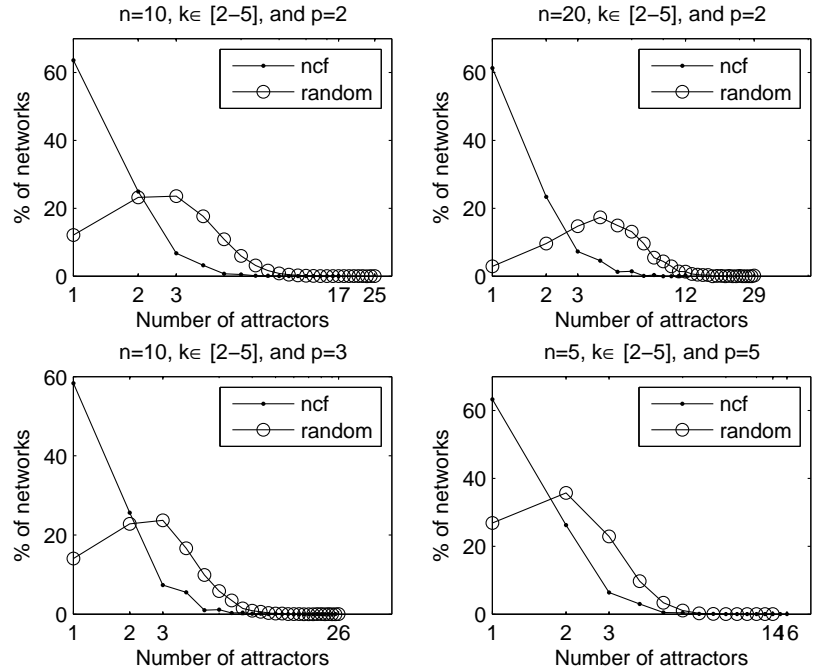}
        \caption{Attractor distribution for networks with nested canalyzing functions (solid circles) and networks with random functions (open circles). The figures show the percentage of networks that returned the number of attractors specified in the x-axis. The parameters $n$, $k$ and $p$ correspond to the number of nodes, the range for the in-degree distribution, and the number of states for each node, respectively. The figures for $n=5,10$ and $p=2,5$ were generated for 1000000 networks, the figure for $n=10$ and $p=3$ was generated for 100000 networks, and the figure for $n=20$ and $p=2$ was generated for 10000 networks.}
        \label{fig:AttractDistMultiState}
  \end{center}
\end{figure}

We present our results concerning attractor distributions in Figure~\ref{fig:AttractDistMultiState}. For each histogram, on the $x$-axis we represent the number of attractors, and on the $y$-axis we represent the percentage of networks that show the given number of attractors specified on the $x$-axis.  The parameters $n$, $k$ and $p$ correspond to the number of nodes, the range for the in-degree distribution, and the number of states for each node respectively. For the experiments performed here we have generated the in-degree distribution from a uniform distribution, independently for each node and network realization. Figure~\ref{fig:AttractDistMultiState} shows very clearly that the number of attractors in nested
canalyzing networks is dramatically smaller than for general networks.  Thus, attractor sizes are larger on average
than in general networks, leading to more robust behavior under perturbations. 

\subsubsection{Cycle length distribution for nested canalyzing networks versus random networks}

\begin{figure}
  \begin{center}
        \includegraphics[width=\textwidth]{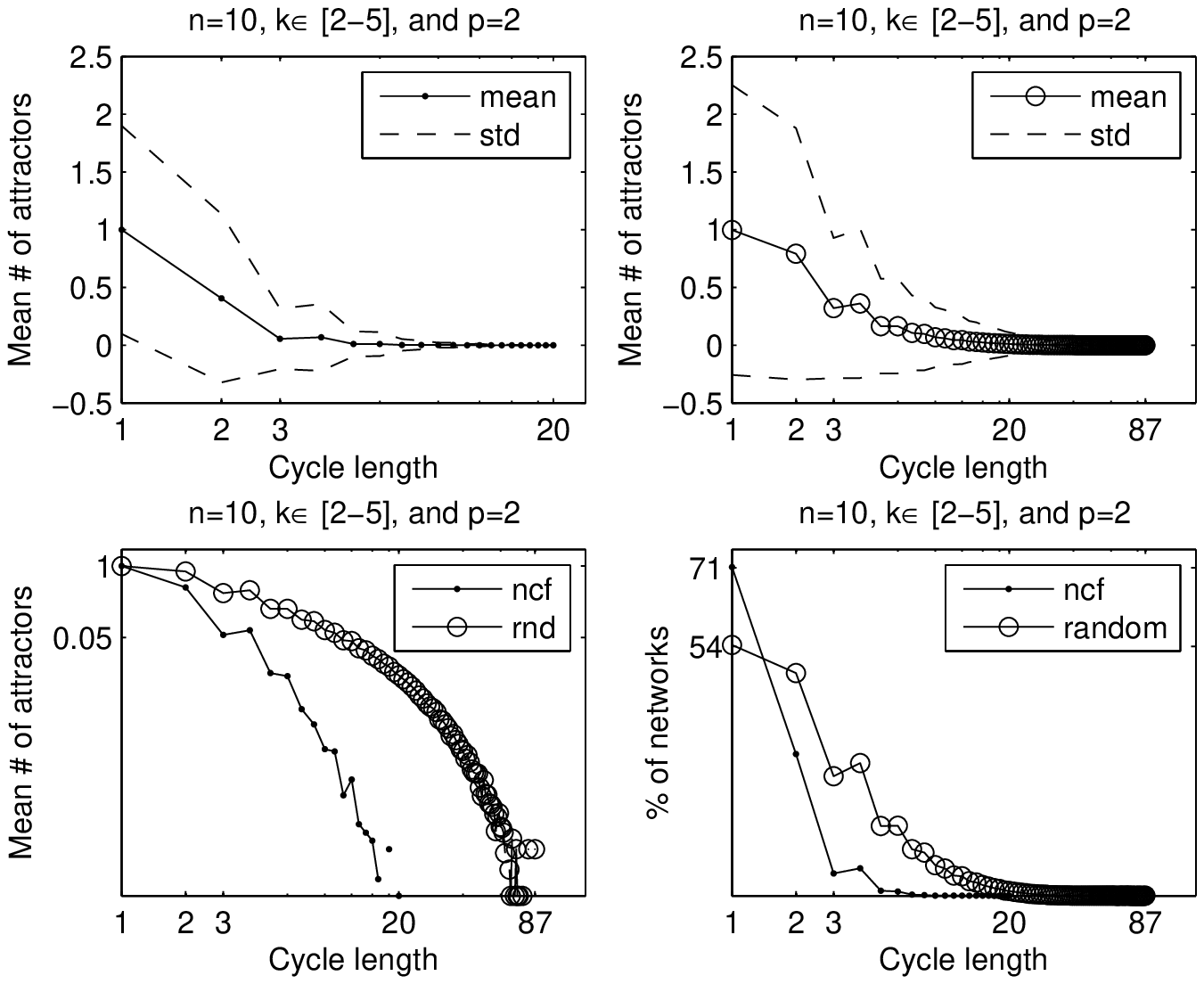}
        \caption{Cycle length for networks with nested canalyzing functions (solid circles) and networks with random functions (open circles). The parameters $n$, $k$ and $p$ correspond to the number of nodes, the range for the in-degree distribution, and the number of states for each node, respectively. The figures were generated for one million networks. The upper figures show the mean number of attractors of length specified in the $x$-axis (solid lines) and their standard deviations (dashed lines), the $x$-axis of these figures are in a logarithmic scale. The bottom left figure shows the mean number of attractors of lengths specified in the $x$-axis in a log-log plot (here rnd means random), and finally the bottom right figure shows the percentage of networks that returned a particular cycle of length specified on the $x$-axis.}
        \label{fig:CycleLength_10_2}
  \end{center}
\end{figure}

\begin{figure}
  \begin{center}
        \includegraphics[width=\textwidth]{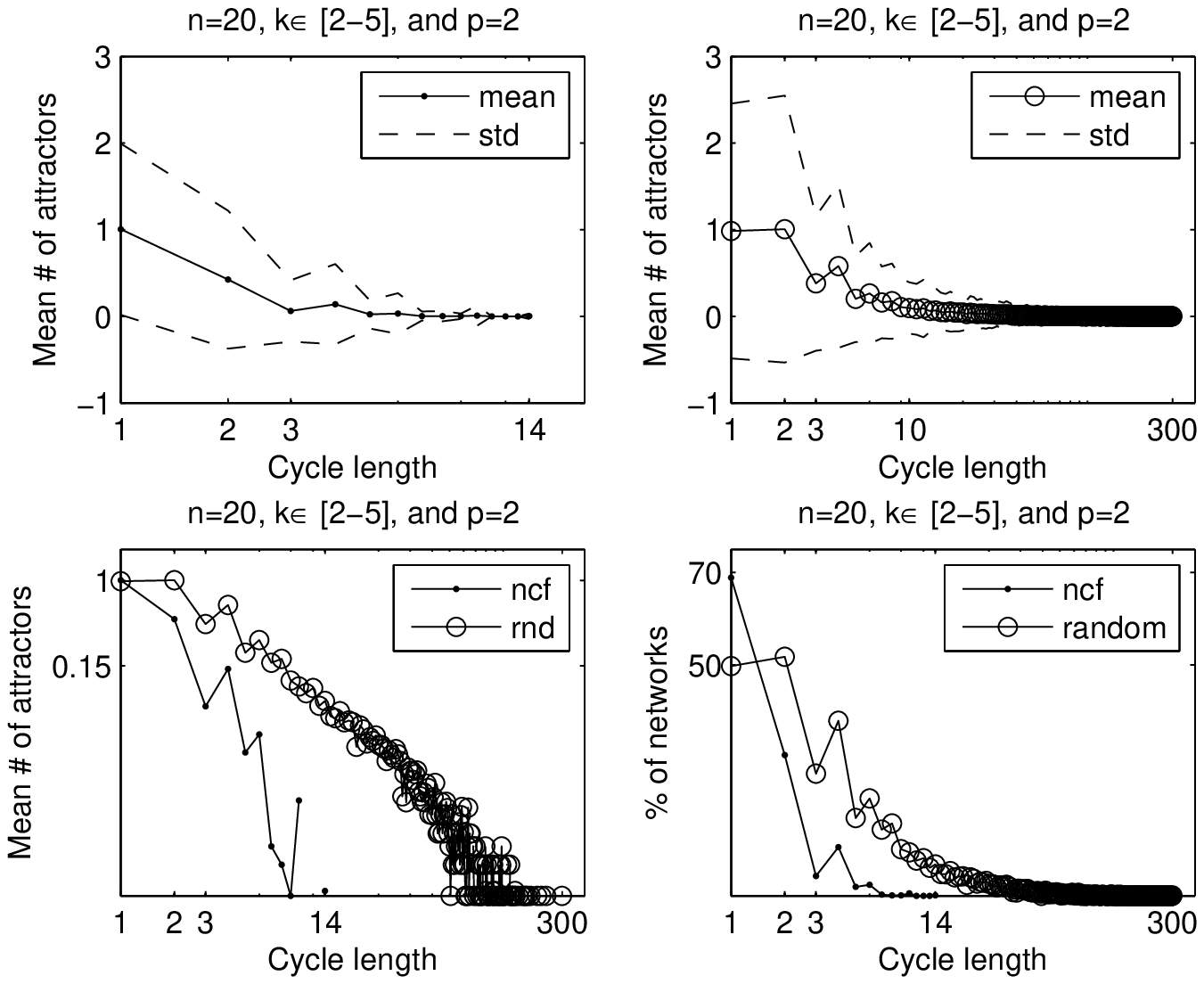}
        \caption{Cycle length for networks with nested canalyzing functions (solid circles) and networks with random functions (open circles). The parameters $n$, $k$ and $p$ correspond to the number of nodes, the range for the in-degree distribution, and the number of states for each node, respectively. The figures were generated for 10000 networks. The upper figures show the mean number of attractors of length specified on the $x$-axis (solid lines) and their standard deviations (dashed lines), the $x$-axis of these figures is a logarithmic scale. The bottom left figure shows the mean number of attractors of lengths specified on the $x$-axis in a log-log plot (here rnd means random), and finally the bottom right figure shows the percentage of networks that returned a particular cycle of length specified on the $x$-axis.}
        \label{fig:CycleLength_20_2}
  \end{center}
\end{figure}

\begin{figure}
  \begin{center}
        \includegraphics[width=\textwidth]{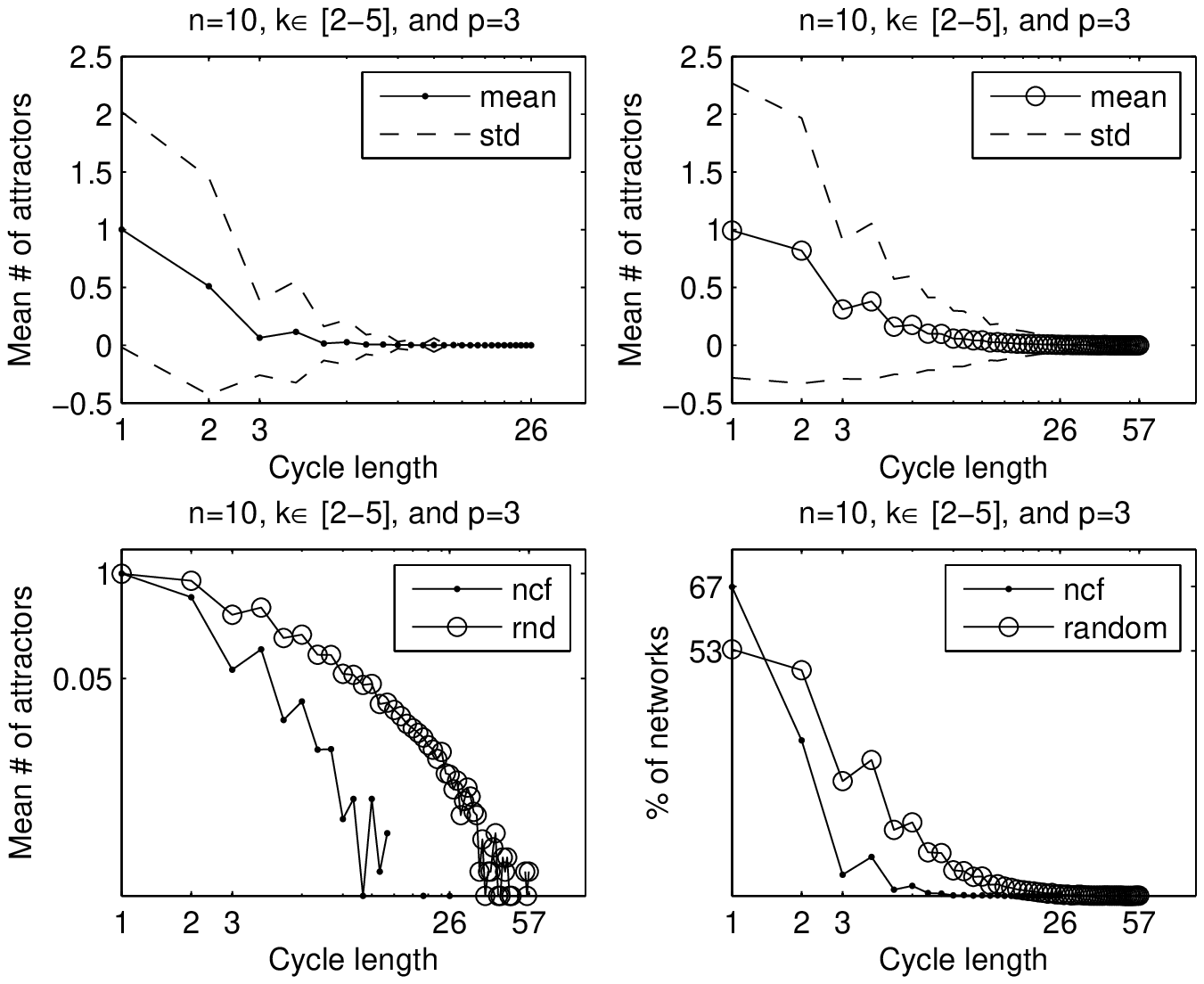}
        \caption{Cycle length for networks with nested canalyzing functions (solid circles) and networks with random functions (open circles). The parameters $n$, $k$ and $p$ correspond to the number of nodes, the range for the in-degree distribution, and the number of states for each node, respectively. The figures were generated for 100000 networks. The upper figures show the mean number of attractors of length specified on the $x$-axis (solid lines) and their standard deviations (dashed lines), the $x$-axis of these figures are in a logarithmic scale. The bottom left figure shows the mean number of attractors of lengths specified on the $x$-axis in a log-log plot (here rnd means random), and finally the bottom right figure shows the percentage of networks that returned a particular cycle of length specified on the $x$-axis.}
        \label{fig:CycleLength_10_3}
  \end{center}
\end{figure}

\begin{figure}
  \begin{center}
        \includegraphics[width=\textwidth]{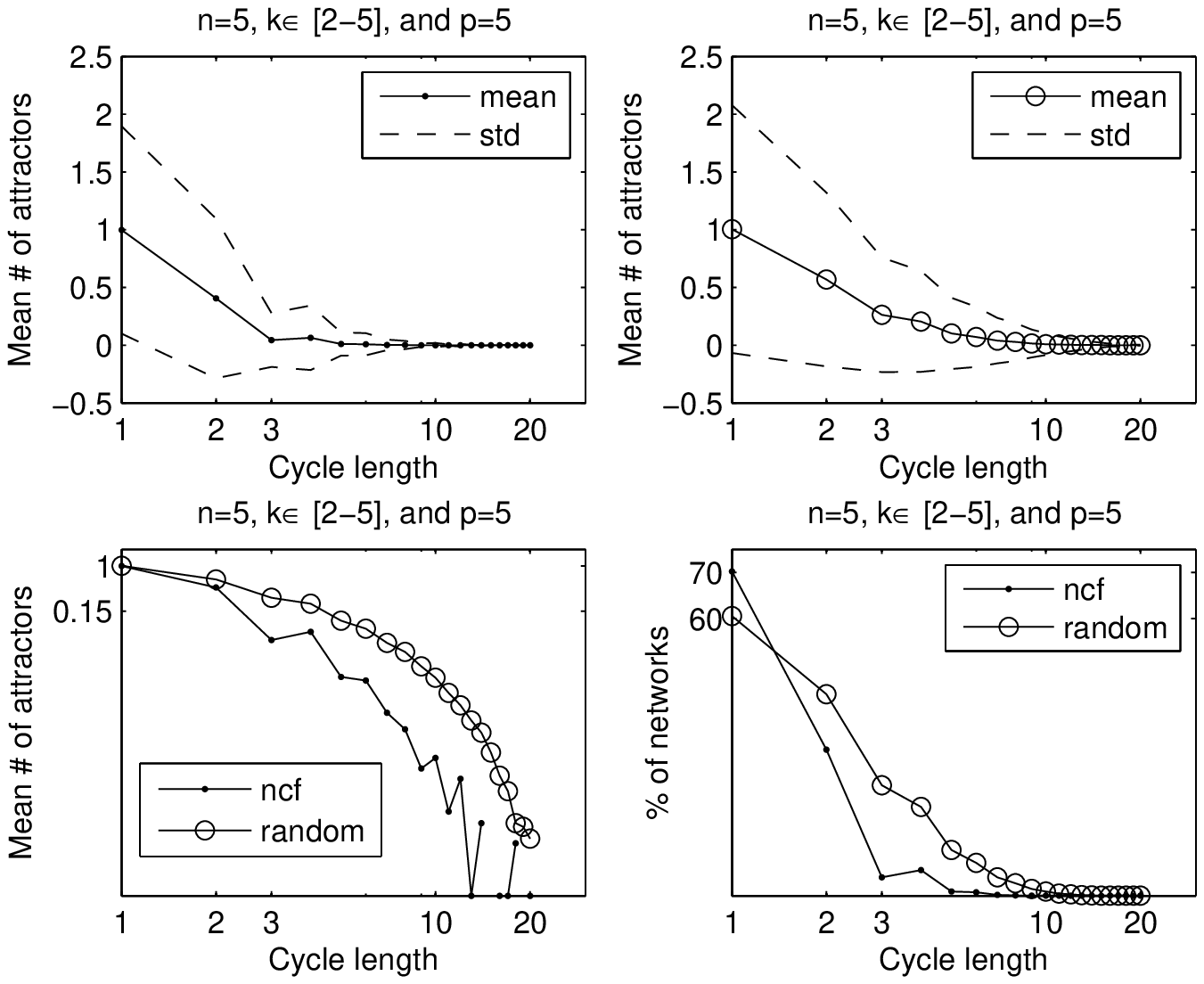}
        \caption{Cycle length for networks with nested canalyzing functions (solid circles) and networks with random functions (open circles). The parameters $n$, $k$ and $p$ correspond to the number of nodes, the range for the in-degree distribution, and the number of states for each node, respectively. The figures were generated for one million networks. The upper figures show the mean number of attractors of length specified on the $x$-axis (solid lines) and their standard deviations (dashed lines), the $x$-axis of these figures are in a logarithmic scale. The bottom left figure shows the mean number of attractors of lengths specified on the $x$-axis in a log-log plot (here rnd means random), and finally the bottom right figure shows the percentage of networks that returned a particular cycle of length specified on the $x$-axis.}
        \label{fig:CycleLength_5_5}
  \end{center}
\end{figure}

We present the results concerning cycle lengths in Figures~\ref{fig:CycleLength_10_2} - \ref{fig:CycleLength_5_5}. For each figure, the upper subfigures show the mean number of attractors of length specified on the $x$-axis (solid lines) and their standard deviations (dashed lines), the $x$-axis of these subfigures are in a logarithmic scale. The bottom left subfigures shows the mean number of attractors of lengths specified on the $x$-axis in a log-log plot, and finally the bottom right subfigures shows the percentage of networks that returned a particular cycle of length specified on the $x$-axis. The parameters $n$, $k$ and $p$ correspond to the number of nodes, the range for the in-degree distribution, and the number of states for each node, respectively. For all of the experiments, we have generated the in-degree distribution from a uniform distribution, independently for each node and network realization.
Figures~\ref{fig:CycleLength_10_2} - \ref{fig:CycleLength_5_5} show clearly that networks with nested canalyzing rules exhibit significantly smaller cycle lengths, leading to more regular behavior. 


\section{Nested canalyzing rules are biologically meaningful}

We hypothesize that nested canalyzing rules are biologically meaningful. To test this hypothesis we have explored 
a range of published multi-state models as to their frequency of appearance. Table~\ref{Table:NCFforLogicalModels} shows that
they are indeed very prevalent, providing evidence that the nested canalyzation is indeed a common pattern for
the regulatory logic in molecular interaction networks. To illustrate this phenomenon we discuss specific examples.
For a complete list of models we have studied see the supporting materials.

\subsection{Examples}

\subsubsection{Lambda Phage Regulation}

\begin{figure}
\begin{center}
\includegraphics[width = 3in]{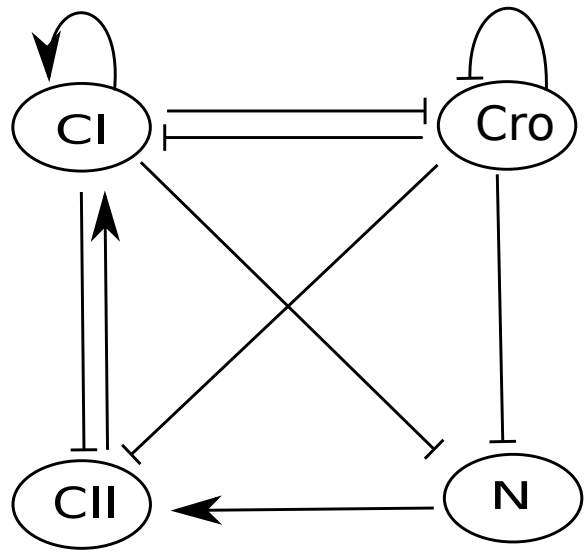}
\caption{Four-variable model for the lambda phage regulatory network.}
\label{LambdaPhage4}
\end{center}
\end{figure}

Thieffry and Thomas~\cite{ThieffryThomas} built a multi-state logical model for the core lambda phage regulatory network. This model encompasses the roles of the regulatory genes \textit{CI}, \textit{CRO}, \textit{CII}, and \textit{N}. See Figure~\ref{LambdaPhage4}.       

The state space for this model is specified by $[0,2]\times[0,3]\times[0,1]\times[0,1]$, that is, the first variable has three levels $\{0,1,2\}$, the second variable has four levels $\{0,1,2,3\}$, and the third and fourth variables are still Boolean.

The update rule for $CI$, $f_{CI}$, has inputs $CRO$ and $CII$ which 
is nested canalyzing in the variable order $CII$, $CRO$, with canalyzing input sets $S_1=\{1\}$, and $S_2=\{1,2,3\}$ 
and canalized output values $2,0,2$, i.e., (see the supporting materials for complete truth tables)

\begin{displaymath}
f_{CI}(CRO,CII)=\left\{
\begin{array}{l}
2 \text{ if } CII \in S_1\\
0\text{ if } CII \notin S_1, CRO \in S_2\\
2\text{ if } CII \notin S_1, CRO \notin S_2.
\end{array}\right.
\end{displaymath}

The update rule for $CRO$, $f_{CRO}$, is nested canalyzing in the variable order $CI$, $CRO$, with canalyzing input sets $S_1=\{2\}$, and $S_2=\{0,1,2\}$, and canalized output values $0,3,2$, i.e.,

\begin{displaymath}
f_{CRO}(CI,CRO)=\left\{
\begin{array}{l}
0 \text{ if } CI \in S_1\\
3\text{ if } CI \notin S_1, CRO \in S_2\\
2\text{ if } CI \notin S_1, CRO \notin S_2.
\end{array}\right.
\end{displaymath}

The update rule for $CII$, $f_{CII}$, is nested canalyzing in the variable order $CII$, $CRO$, $N$, with canalyzing input sets $S_1=\{2\}$, $S_2=\{3\}$, and $S_3=\{1\}$, and canalized output values $0,0,1,0$, i.e.,

\begin{displaymath}
f_{CII}(CI,CRO,N)=\left\{
\begin{array}{l}
0 \text{ if } CI \in S_1\\
0 \text{ if } CI \notin S_1, CRO \in S_2\\
1 \text{ if } CI \notin S_1, CRO \notin S_2, N \in S_3\\
0 \text{ if } CI \notin S_1, CRO \notin S_2, N \notin S_3.
\end{array}\right.
\end{displaymath}

Finally, the update rule for $N$, $f_N$, is nested canalyzing in the variable order $N$, $CRO$, with canalyzing input sets $S_1=\{1,2\}$, and $S_2=\{2,3\}$, and canalized output values $0,0,1$, i.e., 

\begin{displaymath}
f_N(CI,CRO)=\left\{
\begin{array}{l}
0 \text{ if } CI \in S_1\\
0 \text{ if } CI \notin S_1, CRO \in S_2\\
1 \text{ if } CI \notin S_1, CRO \notin S_2.
\end{array}\right.
\end{displaymath}

\subsection{Regulation in the p53-Mdm2 network}

The following model comes from Abou-Jaude W., Ouattara A., Kauffman M.(2009)~\cite{Abou-Jaoude}. The model represents the interactions of the tumor supressor protein p53 and its negative regulator Mdm2.  
Here, P, Mn, Mc, and Dam stand for protein p53, nuclear Mdm2, cytoplasmic Mdm2, and DNA damage, respectively. 
 
The state space for this model is specified by $[0,2]\times[0,1]\times[0,1]\times[0,1]$, that is, except for the first variable $P$ that has three levels $\{0,1,2\}$, all the other variables are still Boolean.
 
\begin{figure}
\begin{center}
\includegraphics[width = 3in]{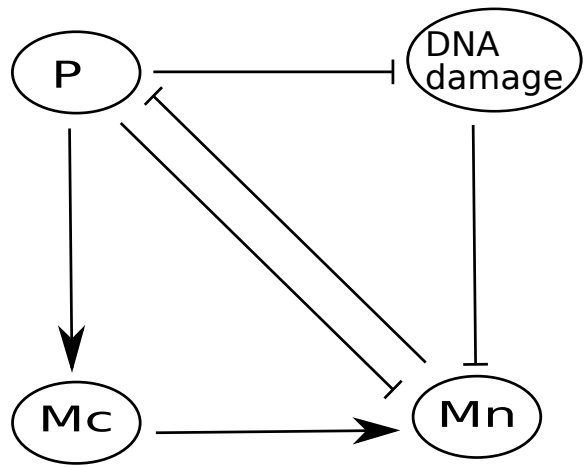}
\caption{Four-variable model for the p53-Mdm2 regulatory network.}
\label{p53WiringDiagram}
\end{center}
\end{figure}

As shown in Figure~\ref{p53WiringDiagram}, $Mn$ acts negatively on $P$. The update rule of $P$, $f_P$, is nested canalyzing with canalyzing input set $S_1=\{0\}$  and canalized output values $K_P, K_{P,\{Mn\}}$, i.e., we can represent $f_P$ as,

\begin{displaymath}
f_P(Mn)=\left\{
\begin{array}{l}
K_p \text{ if}\ Mn \in S_1\\
K_{p,\{Mn\}} \text{ if}\ Mn \notin S_1.
\end{array}\right.
\end{displaymath}
Here $K_P$ is the basal value and $K_{P,\{Mn\}}$ is the parameter value under the influence of $Mn$.

Similarly, for $Mc$, its update rule, $f_{Mc}$, is nested canalyzing with canalyzing input set $S_1=\{0,1\}$ and canalized output values $K_{Mc},K_{Mc,\{P\}}$, i.e., we can represent $f_{Mc}$ as,

\begin{displaymath}
f_{Mc}(P)=\left\{
\begin{array}{l}
K_{Mc} \text{ if}\ P \in S_1\\
K_{Mc,\{P\}} \text{ if}\ P \notin S_1.
\end{array}\right.
\end{displaymath}
Here $K_{Mc}$ is the basal value and $K_{Mc,\{P\}}$ is the parameter value under the influence of $P$. 

For $Mn$, a set of possible parameters for its truth table is given in~\cite{Abou-Jaoude}. We have checked all these cases 
and found that for each case we either get a nested canalyzing function or a constant function. For example, for the second column of Figure 3 (a) in~\cite{Abou-Jaoude} we get that the update rule for $Mn$, $f_{Mn}$, is nested canalyzing in the variable order $Mc$, $P$ with canalyzing input sets $S_1=\{1\}$, $S_2=\{1,2\}$  and canalized output values $1,0,1$, i.e., 
\begin{displaymath}
f_{Mn}(P,Mc)=\left\{
\begin{array}{l}
1 \text{ if}\ Mc \in S_1\\
0 \text{ if}\ Mc \notin S_1, P \in S_2\\
1 \text{ if}\ Mc \notin S_1, P \notin S_2.
\end{array}\right.
\end{displaymath}

When DNA damage is introduced, it has a negative effect on $Mn$. From the set of all possible parameters for its truth table given in inequalities (3)-(5) at~\cite{Abou-Jaoude}, we check that we can always find a nested canalyzing function for its truth table. 
For example, for the third column of Figure 3 (a) in ~\cite{Abou-Jaoude} 
we get that the update rule for Mn (under DNA damage) is nested canalyzing in the variable order $Mc$, $Dam$, $P$, with canalyzing input sets $S_1=\{1\}$, $S_2=\{1\}$, $S_3=\{1,2\}$  and canalized output values $1,0,0,1$, i.e., we can represent $f_{Mn}$ as,

\begin{displaymath}
f_{Mn}(P,Mc,Dam)=\left\{
\begin{array}{l}
1 \text{ if } Mc \in S_1\\
0 \text{ if } Mc \notin S_1, Dam \in S_2\\
0 \text{ if } Mc \notin S_1, Dam \notin S_2, P \in S_3\\
1 \text{ if } Mc \notin S_1, Dam \notin S_2, P \notin S_3.
\end{array}\right.
\end{displaymath}

Finally, the update rule for DNA damage, $f_{Dam}$, is nested canalyzing in the variable order $P$, $Dam$, with canalyzing input sets $S_1=\{2\}$, $S_2=\{1\}$  and canalized output values $0,1,0$, i.e., 
\begin{displaymath}
f_{Dam}(P,Dam)=\left\{
\begin{array}{l}
1 \text{ if } P \in S_1\\
1 \text{ if } P \notin S_1, Dam \in S_2\\
0 \text{ if } P \notin S_1, Dam \notin S_2.
\end{array}\right.
\end{displaymath}



\section{Discussion}
In this paper we have given a definition of a nested canalyzing rule, inspired by the special case of 
Boolean networks, and we have 
shown that it appears as a frequent pattern for the regulatory logic of many molecular interaction networks. 
We have shown that this regulatory pattern leads to networks that have robust and regular dynamics, as
a result of having very small numbers of attractors and very short limit cycles, compared to random networks.
This behavior is also characteristic of many molecular interaction networks. An important application
of this result is to the construction of discrete models, both via a bottom-up or a top-down approach. 
For both approaches, the possibility of restricting the choice of rules to the family of nested canalyzing rules is a 
significant reduction in the possible model space that is available. 

Another interesting aspect of our results is suggested by \cite{Jarrah}. There is was shown that in the
Boolean case, the class of nested canalyzing Boolean rules is in fact identical to the class of 
unate cascade functions. These functions have been studied extensively in computer engineering
and have been shown \cite{Butler} to have the important property that they comprise exactly the class
of Boolean functions that lead to binary decision diagrams with shortest average path length. 
Thus, they make good candidates for the representation of efficient information processing. It would be interesting to study
this property for general nested canalyzing rules. 

The results in \cite{Jarrah} were derived by using a special representation of Boolean functions, namely as
polynomial functions over the Boolean number field, with arithmetic given by addition and multiplication "tables,"
with the key rule that $1+1=0$.  Using a parametrization of the family of all nested canalyzing Boolean polynomials, 
it was shown in \cite{Jarrah}
that the family of all nested canalyzing polynomials in a given number of variables is in fact identical to the
class of unate cascade functions, which, in turn, is equal to the class of Boolean functions that 
result in binary decision diagrams of shortest average path length \cite{Butler}. 
In a paper in preparation we have shown that one can give a similar parameterization of the variety of general nested canalyzing rules, and we 
use this parameterization to derive a formula for the number 
of nested canalyzing rules for a given number
of variables. It would be interesting to investigate whether this general class of nested canalyzing rules 
leads to $n$-ary decision diagrams that have similar properties to those of Boolean nested canalyzing rules.

\clearpage
\begin{table}
  \centering 
    \begin{threeparttable}
\begin{tabular}{| p{7cm} | c | c | c |}
\hline
Model &References & n \tnote{a}& \% NCF\tnote{b} \\
\hline
lysis-lysogeny decision in the lambda phage&\cite{ThieffryThomas}&4&100\%\\
\hline
p53-Mdm2 regulation&\cite{Abou-Jaoude}&4&100\%\\
\hline
Signalling pathways controlling Th cell differentiation&\cite{Naldi}&42&92.8\%\\
\hline
Budding yeast exit module &\cite{Faure,Queralt}&9&77.7\%\\
\hline
Dorsal-ventral boundary formation of the Drosophila wing imaginal disc&\cite{Gonzalez}&24&75\%\\
\hline
Control of Th1/Th2 cell differentiation&\cite{Mendoza,Chaouiya2}&14&71.4\%\\
\hline
Yeast morphogenetic checkpoint&\cite{Faure,Ciliberto}&8&50\%\\
\hline
\end{tabular}
  \caption{Nested canalyzing functions for multi-state models }\label{Table:NCFforLogicalModels}
     \begin{tablenotes}
       \item[a] Number of nodes. Only nodes with in-degree $\geqslant1$ are considered, i.e. non-constant nodes.
       \item[b] Percentage of nodes regulated by nested canalyzing functions.
     \end{tablenotes}
  \end{threeparttable}
\end{table}

\clearpage
\section*{Figure Legends}
%
%

%
\clearpage

\clearpage

\clearpage

\clearpage

\clearpage

\end{document}